# The Kondo effect in magnetic impurities and ferromagnetic contacts

Hyunsoo Yang[1,*], See-Hun Yang[2], Grzegorz Ilnicki[3], Jan Martinek[3], and Stuart S. P. Parkin[2,†]

[1]*Department of Electrical and Computer Engineering, National University of Singapore, 4 Engineering Drive 3, 117576 Singapore*

[2]*IBM Almaden Research Center, 650 Harry Road, San Jose, California 95120, USA*

[3]*Institute of Molecular Physics, Polish Academy of Sciences, Smoluchowskiego 17, 60-179 Poznan, Poland*

## Abstract

Planar macroscopic magnetic tunnel junctions exhibit well defined zero bias anomalies when a thin layer of ferromagnetic CoFe(B) nanodots is inserted within a MgO based tunnel barrier. The conductance curves exhibit a single and a double peak, respectively, for anti-parallel and parallel alignment of the magnetizations of the electrodes which sandwich the tunnel barrier. This leads to a suppression of the tunneling magnetoresistance near zero bias. We show that the double peak structure indicates that the zero-bias anomaly is spin-split due to a magnetic exchange interaction between the magnetic nanodots and the ferromagnetic electrodes. Using a model based on an Anderson quantum dot coupled to ferromagnetic leads, we show that these results imply the coexistence of a Kondo effect and ferromagnetism.

# I. INTRODUCTION

The zero-bias anomaly (ZBA), any departure from a smooth voltage dependent conductance of a tunneling device near zero bias, discovered in the sixties in the planar tunnel junctions doped with paramagnetic impurities became a prototype system to study the Kondo assistance tunneling[1]. It was proposed that a ZBA in the planar tunnel junctions arises from the exchange scattering of conduction electrons by localized paramagnetic states that enhances the tunneling conductance due to the formation of the Kondo resonance in the density of states of impurities at the Fermi level[1,2]. It took three decades of technology development to repeat this experiment in many low dimensional systems including semiconducting quantum dots or single molecules that can be modeled by an Anderson quantum dot[3,4]. Recently, due to further progress in nanotechnology it became possible to study the Kondo effect in single molecules[5], carbon nanotubes[6], self assembled semiconducting quantum dots[7], and quantum point contacts[8] in presence of the ferromagnetic electrodes.

These experimental results raise new questions about the possible coexistence of two many-body effects, the Kondo effect and ferromagnetism in magnetic tunnel junctions (MTJs), that usually compete; does the ZBA arise in the presence of magnetic impurities? How is the ZBA modified in the presence of ferromagnetism? By incorporating metallic, magnetic nanodots within the tunnel barrier, it is expected that MTJs can provide unique Kondo behaviors as compared to previous systems[5-9], depending, for example, on the nanodot size distribution, possible magnetic interactions between the nanodots, and the exchange coupling of the nanodots to the magnetic electrodes. MTJs with magnetic nanodots in the tunnel barriers have been studied recently, but the ZBA due to the Kondo effect was not observed[10-12].

We show that planar MTJs with magnetic nanodots embedded in MgO tunnel barriers can exhibit various ZBAs, including a double-peak (splitting) or a single peak structure for parallel (P) or antiparallel (AP) alignment, respectively, of the ferromagnetic electrodes.

Furthermore, combinations of these two structures are observed depending on the magnetic properties of the nanodots. We find that Kondo physics well accounts for our observations including their dependence on magnetic field and temperature. We report the coexistence of the two many-body effects, namely the Kondo effect and ferromagnetism when a thin layer of magnetic nanodots is inserted within a MgO based MTJ. The Kondo ZBA in our samples is indicative of a strong coupling between the ferromagnetic electrodes and the nanodots through the highly oriented MgO(100) tunnel barriers.

## II. EXPERIMENTAL DETAILS

The junctions were deposited using magnetron sputtering at ambient temperature and were patterned using a sequence of *in-situ* metal shadow masks. The junction area is 700 $\mu$m×700 $\mu$m. The MTJs without nanodots were fabricated with an exchange biased lower ferromagnetic electrode of CoFe (100Å Ta/250Å $Ir_{22}Mn_{78}$/35Å $Co_{70}Fe_{30}$), an upper CoFe counter electrode (70Å CoFe/100Å Ta), and a 28Å MgO tunnel barrier on top of a 8Å Mg layer. The Mg underlayer is used in order to prevent the oxidation of the lower ferromagnetic electrode. MgO barriers are formed by reactive magnetron sputtering in an Ar (98.5%)-$O_2$(1.5%) mixture[13]. MTJs with a nanodot layer were formed from 100 Ta/250 $Ir_{22}Mn_{78}$/35 $Co_{70}Fe_{30}$/8 Mg/ $t$ MgO/$i$ CoFe(B) /8 Mg/ $t$ MgO/70 $Co_{70}Fe_{30}$/150 $Ir_{22}Mn_{78}$/50 Ta, (thicknesses in Å), where $t$ and $i$ denote the thicknesses of the MgO layers and the nominal thickness of nanodot layer, respectively. The exchange bias for the bottom electrode is stronger than that of the top electrode, so that the magnetic moment of each electrode could be independently oriented. A thin $Co_{70}Fe_{30}$ or $Co_{60}Fe_{25}B_{15}$ layer of nominal thickness $i$, inserted in the middle of the MgO layer, forms a discontinuous layer of nanodots when $i <\sim 10$ Å. Electron energy loss spectroscopy reveals that the chemical state of the nanodots is metallic Co and Fe.

The diameter of the nanodots was estimated from transmission electron microscopy

(TEM) images in Fig. 1(a-b). The distribution is described by a log-normal function of the form $f_D(d) = N / (\sigma d \sqrt{2\pi}) \exp[-\ln^2(d/d_m)/(2\sigma^2)]$ whose mean is $d_m \sim 14.8$ Å and standard deviation ($\sigma$) is 0.2 for the case of $i = 2.7$ Å. The average diameter of the nanodots, $d_{avg} = \frac{1}{n} \sum_{i=1}^{n} d_i f_D(d_i)$, is ~16 ± 4 Å. For the case of $i = 8$ Å, the average diameter of the nanodots was estimated to be ~32 ± 7 Å. The magnetic properties of nanodot multilayers of the form [MgO / 5 Å CoFe]$_{20}$ were characterized with a superconducting quantum interference device (SQUID) magnetometer, which shows a typical ferromagnetic loop at 10 K as shown in Fig. 1(c). The mean diameter of the nanodots is estimated to be ~26 Å when $i = 5$ Å.[14] We can estimate the number of dots in our junctions to be of the order of $10^8$. It should be noted that in other material systems in which tunneling through a layer of nanodots has been studied, it has often been observed that the tunneling current is dominated by only a small fraction of the dots. For example, it has been found that the conductance is dominated by a small number of quantum dots (QDs) among $10^6$ to $10^7$ dots in semiconductor heterostructures[15]. We have studied the temperature dependent magneto-conductance characteristics of hundreds of shadow-masked MTJs using standard four-probe methods. Samples with nominally identical structures exhibit similar behaviors.

## III. TRANSPORT RESULTS

### A. Magnetoresistance

Figure 2 plots the dc resistances, R$_P$ and R$_{AP}$, corresponding to P and AP alignment of the ferromagnetic electrodes, respectively, and the resulting TMR=(R$_{AP}$-R$_P$)/R$_P$, for a wide range of temperature and bias voltage (for an MTJ with $t = 24$ Å and $i = 5$ Å). Due to a Coulomb blockade (CB) effect, as the temperature and bias voltage are reduced below a characteristic temperature $T_{CB} = 100$ K down to 4 K, and the CB charging energy $U = 60$ meV, the junction resistance increases at a greater rate, as previously reported in MTJs with a discontinuous layer of nanodots in the tunnel barrier[10-12]. In these previous studies[10-12], due to the weak coupling regime, the transport was dominated by sequential or cotunneling processes, and an increase of the TMR was observed in the CB regime. However, our results, by contrast, show a strong suppression of TMR in the CB regime as shown in Fig. 2(c), which is an important consequence of the Kondo effect for QDs coupled to ferromagnetic electrodes[16-18]. Theory predicts that for the P alignment, due to an exchange interaction between the localized moments and ferromagnetic leads, there is a splitting of the zero-bias anomaly and a reduction of the conductance close to zero-bias. For the AP alignment the exchange interaction from both ferromagnetic electrodes can be compensated, at least partially, which leads to a corresponding enhancement of the conductance and, in consequence, a reduction of TMR[16-18]. This mechanism also explains a strongly non-monotonic dependence of the TMR on the bias voltage as shown in Fig. 2(c).

## B. Double peak structure

Unlike other typical tunneling Kondo experiments which use non-magnetic electrodes and where the conductance peak is at zero bias[3, 4], we observe a double-peak structure at low bias voltage and low magnetic field when the temperature further decreases to 0.25 K, as shown in Fig. 2(a), for $t = 26$ Å and $i = 5$ Å. Even though ferromagnetism is expected to suppress Kondo assisted tunneling[16, 17], we observe a double peak in conductance at low bias voltages and temperatures, and a strong suppression of TMR in the same bias voltage and temperature regime. These features are strongly reminiscent of Kondo resonance effects previously observed in $C_{60}$ molecules, as well as carbon nanotubes, and semiconducting dots placed between ferromagnetic electrodes[5-7]. The splitting of the conductance peaks is too large to be accounted for by Zeeman splitting due to a local magnetic field. Assuming a $g$-factor, $g \approx 2$, a splitting of $\Delta V = 3$ mV corresponds to a magnetic field of 14 T. This large Kondo splitting must therefore be a result of an exchange interaction between the magnetic dots and the ferromagnetic leads, similar to single QD experiments[5, 6]. Because of quantum charge fluctuations, the spin asymmetry in the coupling to the electrodes produces a spin-dependent renormalization of the dot's levels $\varepsilon_\sigma$, breaking the spin degeneracy: $\varepsilon_\uparrow \neq \varepsilon_\downarrow$. This leads to a level spin splitting, $\Delta\varepsilon \equiv \varepsilon_\uparrow - \varepsilon_\downarrow$, that results in a splitting of the ZBA of the conductance curve similar to the effect of an applied magnetic field.

The double-peak structure is hardly observed in samples with thinner MgO barriers ($t$ = 24 Å and $i$ = 5 Å) due to the much steeper increase of the conductance for $|V| > 1.5$ mV in Fig. 4(b), resulting in a single dip. After subtracting a linear background fitted from 5 to 9 mV, the enhanced conductance ($\Delta G$) shows the double peak feature as shown in the inset of Fig. 4(b). For reference, a sample with a thick nanodot layer (e.g. $i$ = 13 Å) does not show any conductance peak or double peaked structure as shown in the inset of Fig. 3(a). As the CoFe nanodot layer is thickened and becomes continuous, the magnetic interaction between the nanodots increases and such a magnetically ordered system completely suppresses the Kondo effect.

## C. Temperature dependence of the linear conductance

As shown in Fig. 3(c), for the sample with $t$ = 26 Å, the peaks of the double-peaked structure gradually overlap as the temperature increases. Indeed, the ZBA completely disappears above $T$ = 15 K, which thus corresponds to the Kondo temperature $T_K$ for this sample. The conductance curves in Fig. 3(c) at each temperature for voltages from 5 to 10 mV are fitted with a second-order polynomial and the fitted background conductance ($G_B$) is calculated at zero bias as shown in Fig. 3(e). $G_B$ is subtracted from the measured conductance data ($G_K$) at zero bias to calculate the enhanced conductance ($\Delta G = G_K - G_B$) at zero bias. The temperature dependence of $\Delta G$ is plotted in Fig. 3(f). In agreement with the theory[19], a plateau in the conductance is found below 2 K, a peak is observed at 8 K, and there is a monotonic decrease above 8 K. For the case of other samples which do not show any exchange splitting, the temperature dependence of the conductance peak can be fitted by the Goldharber-Gordon's empirical Kondo formula[20],

$$\Delta G(T) = G_0 \left[ \frac{\left(T_K^{'}\right)^2}{T^2 + \left(T_K^{'}\right)^2} \right]^s, \text{ where } T_K^{'} = T_K / \sqrt{2^{1/s} - 1} \text{ , } T_K \text{ is the Kondo temperature, and } s \text{ is a}$$

fitting parameter. For $i = 2$ Å, $T_K = 70.8$ K ($s = 4.39$) for the AP state, as shown in the inset of Fig. 3(f). An excellent fitting by this empirical formula is a good supporting evidence for the identification of the zero bias conductance peak as the Kondo peak. A large $s$ value, much bigger than that of a conventional spin 1/2 Kondo system ($s = 0.22$), is attributed to the magnetic moments $> 50$ $\mu_B$ of CoFe(B) nanodots, where $\mu_B$ is the Bohr magneton. For comparison we present calculated plots of $dI/dV$ in Fig. 3(d) using the standard equation-of-motion (EOM) technique. We assume tunneling spin polarization values $P_L = P_R = 0.5$, corresponding to CoFe contacts with an MgO barrier[21] for the left and right electrodes. These polarization values are typical for MgO based MTJs deposited at ambient temperature and not subject to any subsequent anneal treatments[22].

### D. Magnetic field effect

The bias voltage dependence of the conductance of conventional MTJs without nanodots does not show any significant change as a function of applied field if the configuration of the MTJ remains in the same P state shown in Fig. 4(a) for fields ranging from 2 to 8 T. Note that a dramatic conductance change from 0 to 2 T is due to the change of electrodes alignment from AP to P. By contrast, for MTJs with nanodots in the tunnel barrier, we find that the conductance dip widens by twice $g\mu_B B$ with increasing fields, assuming a $g$-factor, $g \approx 2$. Such a broadening, as shown in Fig. 3(b) and 4(c), is one of the distinctive signatures of a spin related effect. To account for this effect we introduce a model that considers a superposition of a Kondo assisted tunneling process (strong coupling) for strongly coupled dots that has a week field dependence on applied field, and a cotunneling process (weak coupling) for weakly coupled dots that shows a stronger field dependence. We model the broadening of the conductance dip by a second-order tunneling process through a single quantum dot coupled to ferromagnetic leads, i.e. cotunneling, that is the simplest possible many-body phenomenon (only two electrons are involved). Using second-order perturbation theory[23, 24] we determine the rate $\gamma_{rr'}^{\sigma\sigma'}$ for a cotunneling process, in which one electron with spin $\sigma'$ tunnels from the dot to the electrode $r'$ while a second electron with spin $\sigma$ enters from the electrode $r$ with the initial and final dot state being $\varepsilon_{\sigma'}$ and $\varepsilon_{\sigma}$, respectively. Here, $r$ = L, R and $\sigma$ = ↑, ↓. For $\sigma = \sigma'$, when the dot spin state is not changed, we can define the elastic cotunneling, $\gamma_{rr'}^{\sigma\sigma}$, while for $\sigma \neq \sigma' = \bar{\sigma}$, (where $\bar{\sigma}$ means opposite spin to $\sigma$), when the dot spin is flipped due to the tunneling process, we can define the inelastic spin-flip cotunneling $\gamma_{rr'}^{\sigma\bar{\sigma}}$ [25]. The elastic cotunneling $\gamma_{rr'}^{\sigma\sigma}$ is insensitive to spin splitting, $\Delta\varepsilon \equiv \varepsilon\!\uparrow - \varepsilon\!\downarrow$, while the inelastic spin-flip cotunneling $\gamma_{rr'}^{\sigma\bar{\sigma}}$ is possible only if the applied bias voltage is larger than the energy difference, $|eV| > |\Delta\varepsilon|$. A linear conductance background has been added to the calculated conductance curve since the conductance curve can be well approximated as linear in the small bias region[26]. A step in conductance in Fig. 4(d), which well mimics that of the experimental data, shifts to higher bias voltage with increasing magnetic field by twice $g\mu_\mathrm{B}B$. This characteristic step corresponds to the voltage at which an inelastic spin-flip cotunneling process is switched on. This occurs when $|eV| > |\Delta\varepsilon|$, where $\Delta\varepsilon = g\mu_\mathrm{B}B$. We find that the best fit to our data is obtained if we add an additional residual field $B_\mathrm{res} = 0.35$ T to the external magnetic field $B$, and if we include strong spin-flip relaxation in the dot[23]. The experimental data in Fig. 4(c) show excellent agreement with the fitted results in Fig. 4(d) based on this model.

## E. The peak splitting dependence on the barrier thickness

The magneto-transport properties of the MTJs are considerably affected by changing the thickness of MgO barriers, as shown in Fig. 5. The splitting of the zero bias anomaly in the conductance data is clearly observable when the thickness of the barrier is relatively thin (28 Å) in Fig. 5(a). For the case of $x = 32$ Å in Fig. 5(b), the amplitude of the zero-bias anomaly becomes very low due to a low value of $T_K$ leading to a flat $dI/dV$ feature around zero bias. As the barrier thickness further increases above 36 Å in Fig. 5(c), the zero-bias anomaly disappears. These data demonstrate that both the peak splitting and the amplitude of the zero bias anomaly (via the value of the Kondo temperature $T_K$) depend on the coupling energy between the localized states and the leads.

## F. Single peak from CoFeB dots

A sample with diluted magnetic nanodots ($Co_{60}Fe_{25}B_{15}$) shows a slightly different type of ZBA as shown in Fig. 6(a). A single broad conductance peak is observed at zero bias in both the AP and P configurations but in the P configuration an incipient double peak structure can be seen. The peak separation $\Delta V = 10$ mV corresponds to more than a 50 T field, if $g \approx 2$. This observation can also be accounted for by taking into account Kondo assisted tunneling and the exchange interaction between the nanodots and the ferromagnetic leads. Using the EOM technique used above with $U = 100$ meV, $P = 0.5$, $\Gamma = 10$ meV, and $\varepsilon_0 = -4\Gamma$, qualitative agreement with the experiment is found, as shown in Fig. 6(b), although the measured peak width at zero bias for the AP configuration is substantially broader that the calculated value. This discrepancy in widths is attributed to an incomplete compensation of exchange fields that may be due to an asymmetric coupling with the left and right leads ($\Gamma_L \neq \Gamma_R$). This is not unlikely since it is very difficult to prepare two identical MgO tunnel barriers and interfaces with the magnetic elements. We can estimate the magnitude of the splitting as $e\Delta V = 2a\sum_{r=L,R} P_r\Gamma_r$, where $a$ is a constant of order unity, whose magnitude and sign depend on the charging energy and the details of the band structure[5, 17]. In the AP configuration, $e\Delta V_{AP} = 2aP(\Gamma_L\text{-}\Gamma_R)$ since $P_L = -P_R = P$. Therefore, $\Delta V_{AP} = 0$ only when $\Gamma_L = \Gamma_R$, which thereby accounts for the broad conductance peak, due to incomplete compensation in Fig. 6(a). Due to the dilution of nonmagnetic material the interaction between nanodots is reduced, and therefore the system is closer to the ideal Anderson model.

## IV. DISCUSSIONS

In our samples we are not able to observe a CB staircase phenomenon. Firstly, as we clearly show in our TEM images (see Fig. 1), our devices contain distributions of nanodots of various sizes and shapes, so that the Coulomb energies of each dot will be different and thus any staircase would be smeared out. Secondly and more importantly, typically CB staircase phenomena are observed when a gate voltage is applied to the particle or nanodot. Our devices are planar MTJs in which there is no gate electrode with which we could apply a gate voltage to the nanodots, and, moreover, there is a very large value of the charging energy which is typical for planar tunnel junctions, that requires a very large voltage to bring their electronic states into resonance with the Fermi level of the electrodes. In nearly all other experiments a significant gate bias voltage has to be applied in order to observe a CB staircase. Also in non-magnetic tunnel junction experiments where the ZBA due to the Kondo effect was studied[27-29], the CB staircase was not observed. However, the results published in all these papers are well established evidence of the Kondo effect even without detection of the CB staircase. In addition, a recent paper by H. Sukegawa et al.[11] that discussed tunneling through nanodots in the CB regime, and whose data are similar to ours in the same regime of nanoparticle size i.e. for cotunneling could observe no CB steps.

There are several important reasons why the magnitude of the effect that we attribute to Kondo assisted tunneling will much smaller than that in the unitary limit. Even though the nominal growth thicknesses of the two tunnel barriers are designed to be the same, this is unlikely in practice because the lower barrier is grown on a flat CoFe layer whereas the upper

MgO barrier is grown on the layer of CoFe nanodots which is of a non-uniform thickness. The tunneling characteristics of the two barriers is consequently very unlikely to be identical which will likely lead to much smaller conductance values well below the unitary limit. For example, a factor of more than a 100 fold reduction in conductance was reported in recent experiments on tunneling through single molecular quantum dots.[30]

A more important reason why we would not expect the conductance of our devices to be simply the conductance through each dot in the unitary limit multiplied by the number of channels (i.e. dots) is that only a small fraction of the dots is likely to be in resonance at zero bias due to the distribution of sizes and shapes of the nanodots. Typically, in Kondo experiments of tunneling through single quantum dots, such as artificial atoms in GaAs or break junctions with single molecular quantum dots, a gate electrode is used to apply a voltage to the quantum dot in order to bring the energy of the corresponding electronic state on the dot into coincidence with the Fermi energy of the electrodes in order to observe a Kondo tunneling conductance peak. In one recent experiment which studied Kondo tunneling through nominally identical single C60 molecules, without a gate bias, a Kondo peak was reported in only a small fraction (~15%) of the devices fabricated[30]. In nearly all other experiments a significant gate bias voltage has to be applied in order to observe a Kondo conductance peak.

Perhaps it is interesting to point out that in other material systems in which tunneling through a layer of nanodots has been studied, it has often been observed that the tunneling current is dominated by only a small fraction of the dots. For example. Gould et al. report on tunneling through self assembled CdSe quantum dots in a tunnel barrier formed from ZnBeMnSe[15]. They and works by other groups on related semiconductor QD structures [31, 32] find that the conductance is dominated by a small number of QDs even though there are 6 or 7 orders of magnitude greater number of dots in the barrier.

Finally, the distribution in the alignment of the nanoparticle's magnetic moments with respect to the electrodes' moments clearly influences the magnitude of the conductance through them and this is also likely to be a significant factor in reducing the magnitude of their conductance. We also point out that there are many observations in the literature in which it has been observed that small changes in the geometry of tunneling devices can strongly reduce the magnitude of the conductance from the unitary limit. Thus, the conductance of our devices, even though much lower than a simpleminded calculation of the conductance in the unitary limit, is not unreasonable and is consistent with other studies on Kondo tunneling through nanoscopic entities in the Coulomb dot regime.

## V. SUMMARY

We have demonstrated that planar macroscopic MTJs with a layer of magnetic nanodots placed inside an MgO tunnel barrier exhibit Kondo assisted tunneling at low bias voltage and temperature with a zero bias conductance Kondo peak anomaly. This peak is split in the absence of any applied field due to the exchange coupling of the nanodots to the ferromagnetic electrodes. In this regime the TMR is suppressed which is evidence of Kondo assisted spin flip tunneling. Our results can be well accounted for within an Anderson model using the spin-1/2 Kondo model of non-magnetic quantum dots coupled to ferromagnetic leads even though the impurities are magnetic. We find that the competition between Kondo assisted tunneling and magnetic exchange coupling of the nanodots to the ferromagnetic leads play key roles in determining the detailed dependence of the tunneling conductance on bias voltage, temperature, and magnetic field.

## ACKNOWLEDGMENTS

This work was supported by the Singapore Ministry of Education Academic Research Fund Tier 2 (MOE2008-T2-1-105), the Singapore NRF CRP Award No. NRF-CRP 4-2008-06, and a Polish grant for science for the years 2009-2012.

**APPENDIX: EOM TECHNIQUE**

In our calculations we use the standard Hamiltonian for the Anderson QD with a single energy level $\varepsilon_0$ coupled to ferromagnetic leads:

$$H = \sum_{kr\sigma} \varepsilon_{kr\sigma} c^+_{rk\sigma} c_{rk\sigma} + \sum_{\sigma} \varepsilon_{0\sigma} d^+_{\sigma} d_{\sigma} + U n_{\uparrow} n_{\downarrow} + \sum_{rk\sigma} (\nu_{r\sigma} d^+_{\sigma} c_{rk\sigma} + h.c.), \tag{A1}$$

where $c_{rk\sigma}$ and $d_{\sigma}$ are Fermi operators for electrons with a wave vector $k$ and spin $\sigma$ in the leads, $r = L, R$, and in the QD, respectively. Here $t_{rk}$ is the tunneling matrix element, and the Zeeman energy of the dot is given by $\Delta\varepsilon = \varepsilon_{\uparrow} - \varepsilon_{\downarrow} = g\mu_B B$. The ferromagnetism of the leads is accounted for by different densities of states (DOS) for up- and down-spin electrons, $\nu_{r\uparrow}(\omega) \neq \nu_{r\downarrow}(\omega)$. Using the Keldysh formalism, the electric current $I = \sum_{\sigma} I_{\sigma}$ through a QD for $\Gamma_{R\sigma}(\omega) = \lambda_{\sigma}\Gamma_{L\sigma}(\omega)$ is given by,

$$I_{\sigma} = \frac{e}{\hbar} \int d\omega \frac{\Gamma_{L\sigma}(\omega)\Gamma_{R\sigma}(\omega)}{\Gamma_{L\sigma}(\omega) + \Gamma_{R\sigma}(\omega)} \left[ f_L(\omega) - f_R(\omega) \right] \rho_{\sigma}(\omega) \tag{A2}$$

where $\rho_{\sigma}(\omega) = -1/\pi \, \mathrm{Im} \, G^{ret}_{\sigma}(\omega)$ and the coupling energy $\Gamma_{r\sigma}(\omega) = 2\pi |t|^2 \nu_{r\sigma}(\omega)$. For strong interaction the retarded Green function can be found as,

$$G^{ret}_{\sigma}(\omega) = \frac{1 - \langle n_{\bar{\sigma}} \rangle}{\omega - \varepsilon_{\sigma} - \Sigma_{0\sigma}(\omega) - \Sigma_{1\sigma}(\omega) + i0^+} \tag{A3}$$

where $\Sigma_{0\sigma}(\omega) = \sum_{k \in L,R} |t_k|^2 / (\omega - \varepsilon_{k\sigma})$ is the self-energy for a noninteracting QD, while

$$\Sigma_{1\sigma}(\omega, \Delta\tilde{\varepsilon}) = \sum_{k \in L,R} \frac{|t_k|^2 f_{L/R}(\varepsilon_{k\sigma})}{\omega - \sigma\Delta\tilde{\varepsilon} - \varepsilon_{k\sigma} + i\hbar/2\tau_{\bar{\sigma}}} \tag{A4}$$

appears for an interacting QD only. The average occupation of the QD with spin $\sigma$ is obtained from $\langle n_\sigma \rangle = -i/2\pi \int d\omega \, G_\sigma^<(\omega)$. Following a reference[17] we replace on the right-hand side of Eq. (A1) $\Delta\varepsilon \to \Delta\tilde{\varepsilon}$, where $\tilde{\varepsilon}_\sigma$ is found self-consistently from the relation,

$$\tilde{\varepsilon}_\sigma = \varepsilon_\sigma + \mathrm{Re}\left[\Sigma_{0\sigma}(\tilde{\varepsilon}_\sigma) + \Sigma_{1\sigma}(\tilde{\varepsilon}_\sigma, \Delta\tilde{\varepsilon})\right] \qquad (A5)$$

which describes the renormalized dot-level energy, where the real part of the denominator of Eq. (A3) vanishes. In the fitting procedure we assume, for simplicity, a flat band structure $\nu_{r\sigma}(\omega) = \nu_{r\sigma}$ and neglect the k-dependence of the tunneling amplitudes, $t_{rk} = t$. In the fitting procedure we use the spin polarization of the coupling energy defined as $P_r \equiv (\Gamma_{r\uparrow} - \Gamma_{r\downarrow})/(\Gamma_{r\uparrow} + \Gamma_{r\downarrow})$ whose value is extracted from previous experiments[21], and as the fitting parameters, the total coupling energy $\Gamma = \frac{1}{2}(\Gamma_L + \Gamma_R)$, (where $\Gamma_r = \Gamma_{r\uparrow} + \Gamma_{r\downarrow}$), the QD energy level $\varepsilon_0$, and the interaction energy $U$.

[*] Electronic address: eleyang@nus.edu.sg
[†] Electronic address: parkin@almaden.ibm.com

Figure. 1: (a,b) Plan view TEMs when a thin CoFe layer of nominal thickness *i* is inserted in the middle of the MgO layer. (c) Magnetization versus magnetic field curve of the form [MgO / 5 Å CoFe]$_{20}$ at 10 K. The inset shows the histogram of nanodot diameters where the solid red line is a fit of a log-normal distribution function.

Figure. 2: Temperature and bias voltage dependence of the dc resistance for P (a) and AP (b) magnetic configurations of a MTJ with a structure: CoFe/24 Å MgO/5 Å CoFe/24 Å MgO/CoFe. (c) TMR of the same junction. The inset is the schematic illustration of a cross-section of the MTJ with a layer of CoFe nanodots within the MgO layer. Magnetic field of 10,000 Oe and -500 Oe was applied to set the state of the MTJs in the P and AP states, respectively.

Figure. 3: (a, b) Magnetic field dependence of the conductance at $T = 0.25$ K for a junction composed of CoFe/26 Å MgO/5 Å CoFe/26 Å MgO/CoFe. (c) Temperature dependence at 0.25, 0.5, 1, 2, 4, 8, 15, 20, and 25 K of the conductance at zero field. (d) Theoretical fit to (c) using the EOM method. The offset was adjusted. The total coupling parameter to the ferromagnetic leads $\Gamma = 1/2(\Gamma_L + \Gamma_R) = 3$ meV, the dot's energy level $\varepsilon_0 = -10$ meV (with respect to the Fermi energy), and $U = 100$ meV, where $\Gamma_L(\Gamma_R)$ denotes coupling to the left (right) electrode. The curves are relatively insensitive to changes in $U$ by up to an order of magnitude. (e) Plot of $G_B$ versus temperature for the data shown in (c). (f) Temperature dependence of the linear conductance enhancement ($\Delta G$) for the data in (c). The inset in (a) shows the conductance from an MTJ of the form: CoFe/28 Å MgO/13 Å CoFe/28 Å MgO/CoFe. The inset in (e) shows how $G_K$ and $G_B$ are determined. The inset in (f) shows the temperature dependence of $\Delta G$ for $i = 2$ Å.

Figure. 4: Conductance versus bias voltage data at $T = 0.25$ K for various magnetic fields for a MTJ device without a nanodot layer (a) and for an MTJ composed of CoFe/24 Å MgO/5 Å CoFe/24 Å MgO/CoFe for a wide (b) and narrow (c) range of bias voltage. (d) Theoretical fit to (c) using a second-order perturbation (cotunneling) theory. The inset in (b) shows the enhanced conductance after subtracting a linear background fitted from 5 to 9 mV at 0 T.

Figure. 5: Differential conductance versus bias voltage curves for various MgO thicknesses at 2.6 K for the P configuration for MTJs comprised of 50 Ta/ 250 $Ir_{22}Mn_{78}$/ 3 $Co_{49}Fe_{21}B_{30}$/ 37 $Co_{70}Fe_{30}$/ 8 Mg/ 25 MgO/ 5 $Co_{70}Fe_{30}$/ 8 Mg/ $x$ MgO/ 70 $Co_{70}Fe_{30}$/ 150 $Ir_{22}Mn_{78}$/ 50 Ta, (thicknesses in Å). Magnetic field of 10,000 Oe was applied to set the state of the MTJs in the P states.

Figure. 6: (a) Bias voltage dependence of conductance at $T$ = 2.6 K in various magnetic fields for a MTJ composed of CoFe/28 Å MgO/5 Å CoFeB/28 Å MgO/CoFe, for P (4553, 1920, 1008, and -4988 Oe) and AP (-101, -500, and -1033 Oe) alignments. (b) Theoretical fit to (a) using the EOM method and taking into account symmetric couplings to the left and right leads $\Gamma_L = \Gamma_R$.

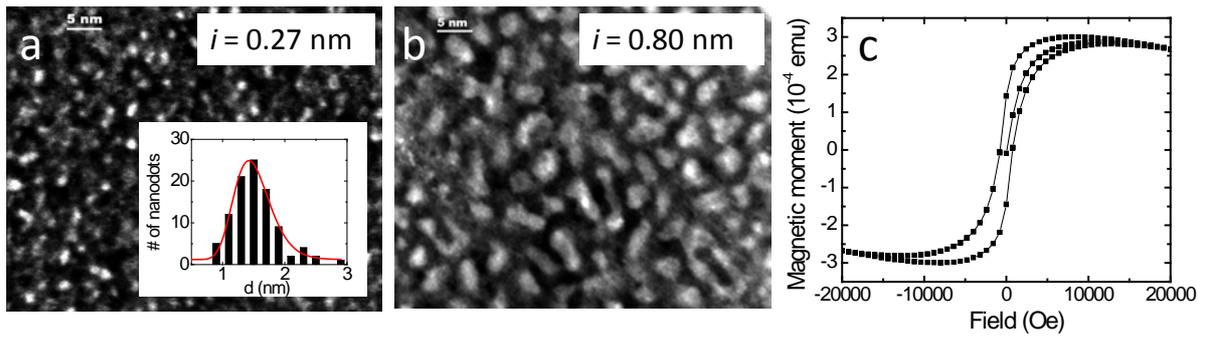

Figure 1

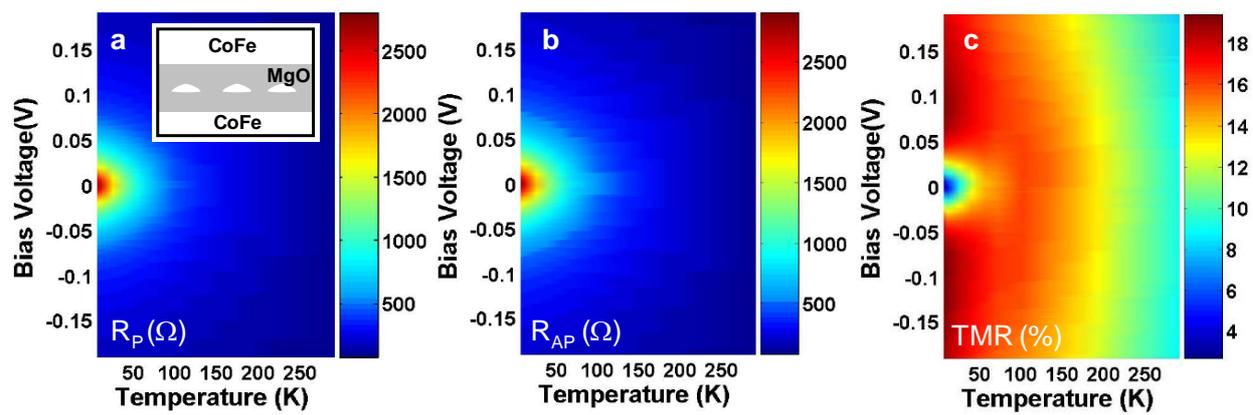

Figure 2

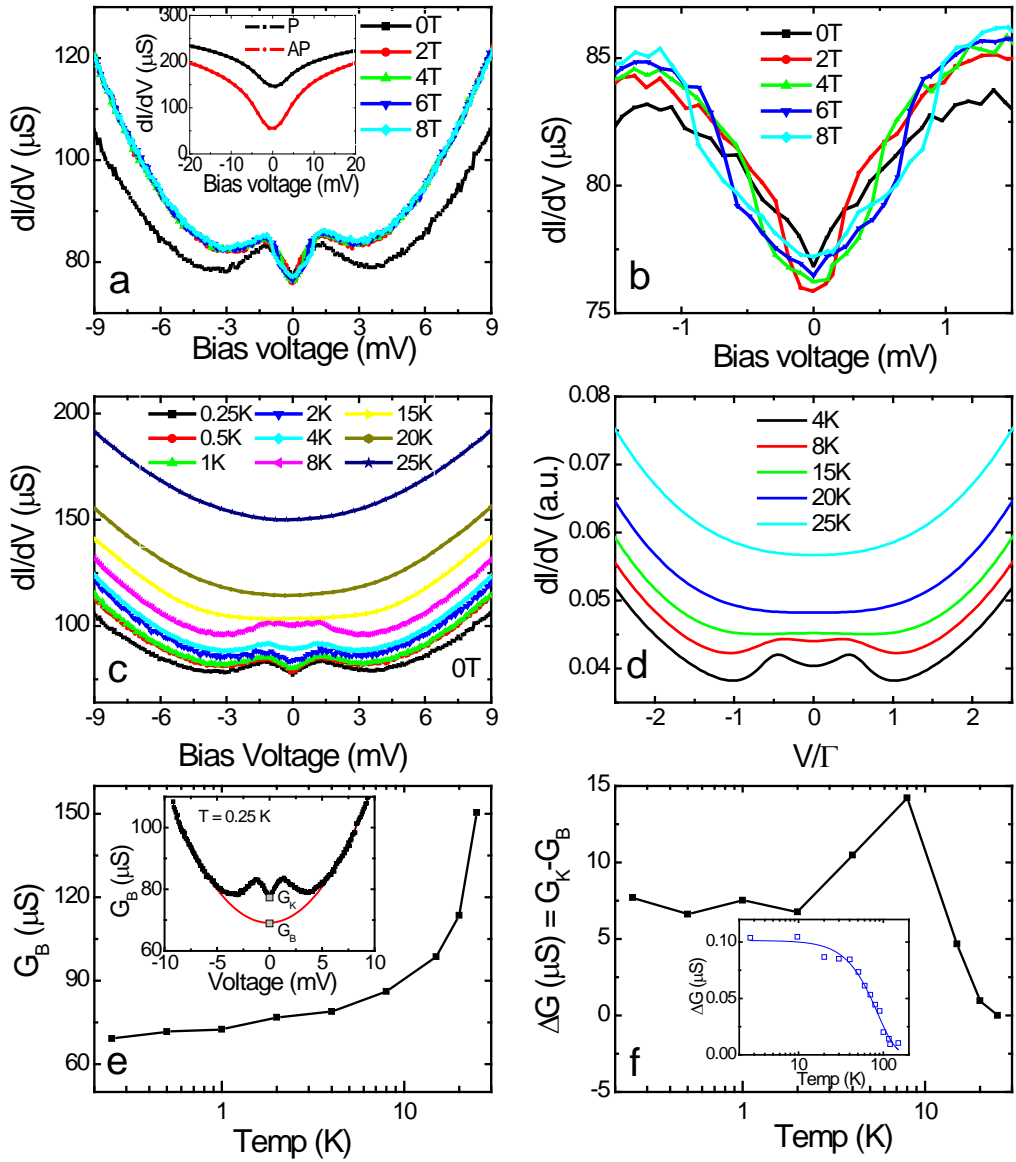

Figure 3

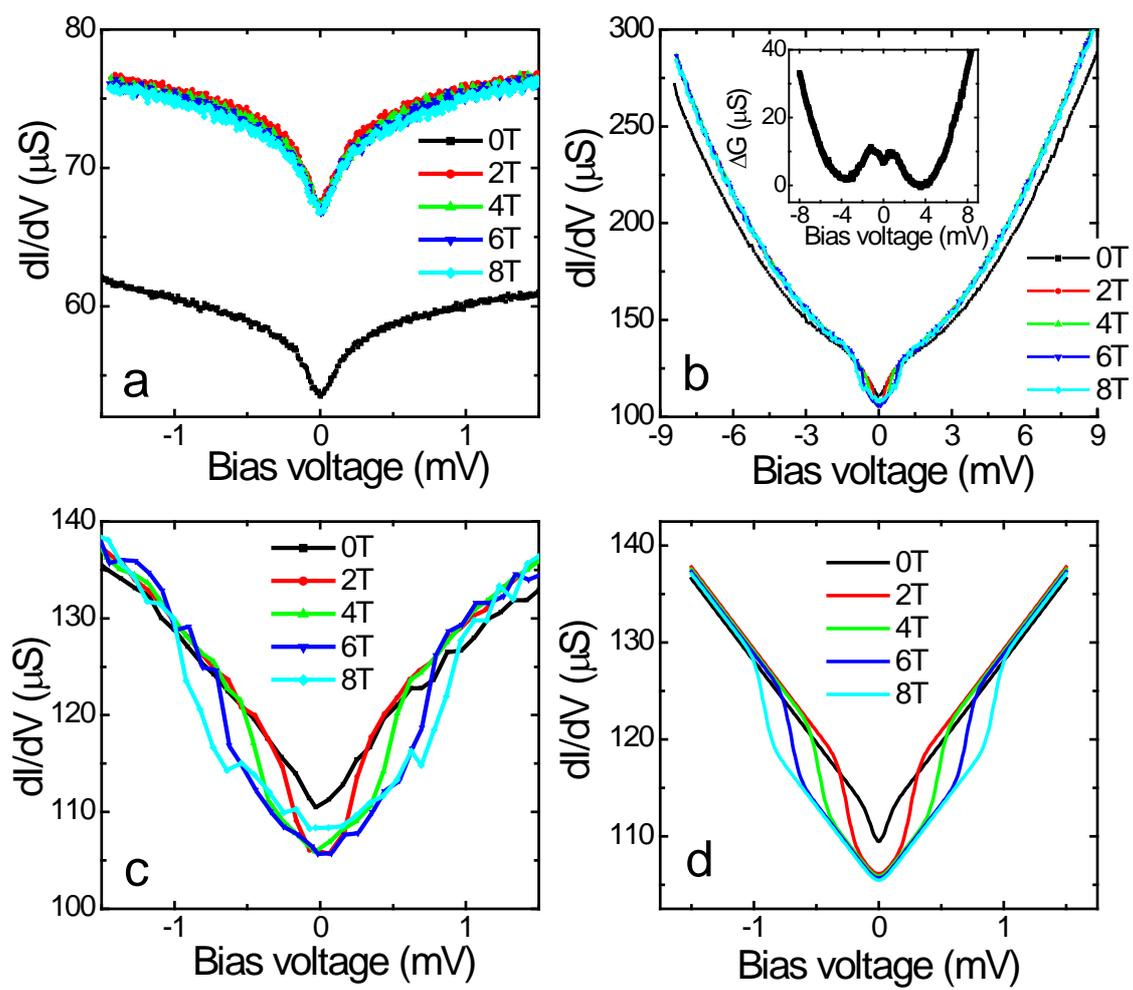

Figure 4

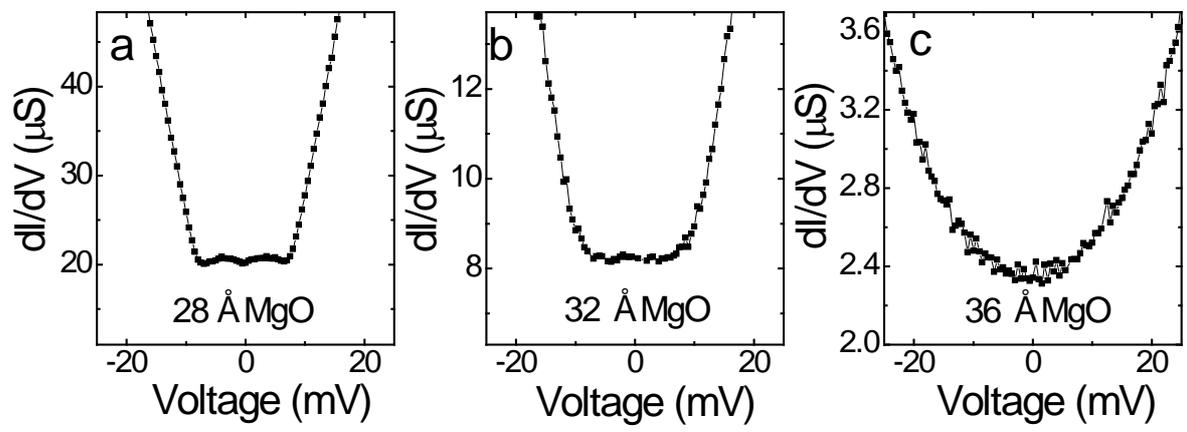

Figure 5

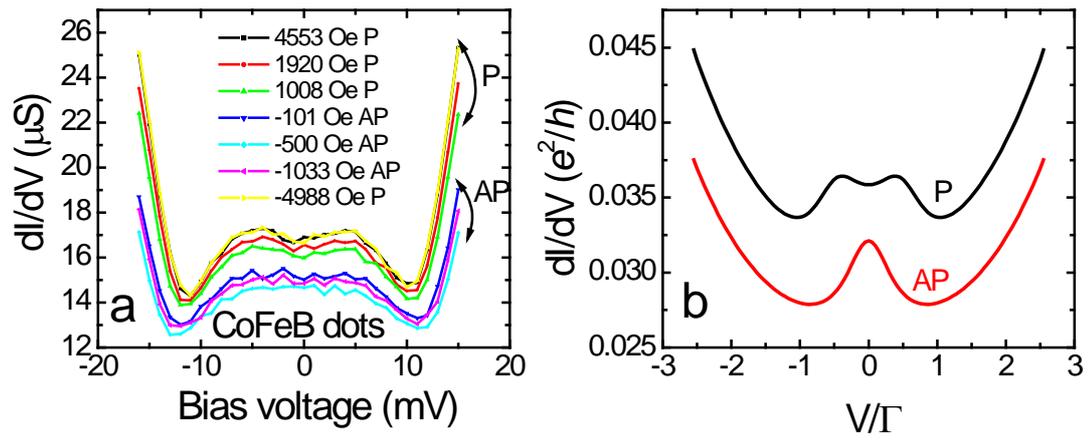

Figure 6